# Modeling Envisat RA-2 waveforms in the coastal zone: case-study of calm water contamination

Gómez-Enri[1], J., S. Vignudelli[2], G.D. Quartly[3], C.P. Gommenginger[3], P. Cipollini[3], P.G. Challenor[3] and J. Benveniste[4]

1 - University of Cadíz, Spain. jesus.gomez@uca.es
2 - Consiglio Nazionale delle Ricerche (CNR), Italy
3 - National Oceanography Centre, Southampton, UK
4 - ESA/ESRIN, Italy

# MODELING ENVISAT RA-2 WAVEFORMS IN THE COASTAL ZONE: CASE-STUDY OF CALM WATER CONTAMINATION

Gómez-Enri, J., S. Vignudelli, G.D. Quartly, C.P. Gommenginger, P. Cipollini, P.G. Challenor and J. Benveniste

**Abstract**

Radar altimeters have so far had limited use in the coastal zone, the area with most societal impact. This is due to both lack of, or insufficient accuracy in the necessary corrections, and more complicated altimeter signals. This paper examines waveform data from the Envisat RA-2 as it passes regularly over Pianosa (a 10 km$^2$ island in the NW Mediterranean). Forty-six repeat passes were analysed, with most showing a reduction in signal upon passing over the island, with weak early returns corresponding to the reflections from land. Intriguingly one third of cases showed an anomalously bright hyperbolic feature. This feature may be due to extremely calm waters in the Golfo della Botte (northern side of the island), but the cause of its intermittency is not clear. The modelling of waveforms in such a complex land/sea environment demonstrates the potential for sea surface height retrievals much closer to the coast than is achieved by routine processing. The long-term development of altimetric records in the coastal zone will not only improve the calibration of altimetric data with coastal tide gauges, but also greatly enhance the study of storm surges and other coastal phenomena.

**1. Introduction**

Radar altimetry is a tool primarily designed for global retrieval of sea surface height (SSH) from space. A complex sequence of processing steps is usually necessary to transform raw data into usable geophysical information. These steps essentially consist of removing unwanted effects caused by the instrument, atmosphere and ocean [1]. The standard products contain sensor measurements, orbit estimations and a full set of corrections, at ~7 km along track, which resolution is normally sufficient for open ocean studies, but is too coarse for many applications in the coastal zone.

Two particular problems exist in the coastal zone. One is the degradation of altimeter corrections (principally for tides and wet troposphere) and the other relates to the performance of the altimeter itself. The resolution of both these issues has been recognised by the space agencies to be of critical importance for the further exploitation of altimetry near land, with the French space agency, Centre National d'Etudes Spatiales (CNES) funding PISTACH (Prototype Innovant de Système de Traitement pour les Applications Côtières et l'Hydrologie) to look at data from the Jason altimeters, and the European Space Agency (ESA) supporting COASTALT to examine the potential improvements to Envisat RA-2 data. Some studies have focussed on improving corrections and data processing [2], [3], but few papers have yet tackled the problem of interpreting altimeter data arising from mixed land/ocean surfaces [4], which is the aspect of the problem we tackle here.

Over a uniformly rough ocean surface the mean radar echo has a well-defined shape, with a steeply rising leading edge followed by a gradual decline in power over the rest of the waveform. This shape, resulting from the convolution of the emitted pulse, the flat surface response and the vertical distribution of surface scatterers, is termed "Brown-like" after one of its original proponents [5]. Typically this model is fitted (in the least-square sense) to the amplitude waveforms to derive the elapsed travel time (related to the range), the maximum amplitude of the signal (related to wind speed at the sea surface) and the variability in surface height denoted by the significant wave height (SWH). [6] and [7] developed the theory for determining another parameter, SSH skewness, which was successfully implemented for Jason-1 [8] and Envisat RA-2 [9]

However pulse echoes are more complex and variable when there is significant spatial variation of properties within the full altimetric footprint i.e. that portion of the surface area contributing to any of the altimeter's waveform sampling gates (a disc ~14 km across for Envisat RA-2). [10] was one of the first to demonstrate the complexity of signals close to the coast. There are two clear effects: first the reflectivity of land will be different from that of the surrounding ocean, and second, the land will provide earlier returns (as it is nearer to the satellite). Here we examine the challenges in interpreting and modelling waveform data near a small island. Section 2 describes the location of the case study, along with the particular data to be used. Section 3 shows the spatial and temporal variation in the collated waveform data, and demonstrates how these effects can be modelled. The fourth section discusses the possible causes of the features found, and the concluding section summarises the results and discusses the global implications.

## 2. Case study and data

**a. Pianosa Island**

Pianosa is a small island about 10 km$^2$ in area in the Tuscan Archipelago between Corsica and Italy (Fig.1). Its name comes from the Italian word "pianura" (plain), which aptly describes the island, practically all flat, reaching a maximum height of 29 m above sea level and with an average height of 10 m. The lack of relief contributes to the scarcity of rain, but some vegetation cover makes it a weak reflector of radar waves. To assess data over this island we use a high-resolution coastal line and digital elevation model (DEM), with a horizontal resolution of 10m.

**b. Waveform data**

Our chief interest in Pianosa is because it lies on Envisat RA-2 repeat ascending track 128, and so there are data from this instrument every 35 days from cycle 11 (November 2002) to the present. The routine altimeter records provide 1.1-second averages (normally called "1Hz" data) of the derived geophysical variables; here we use the "18Hz" Sensor Geophysical Data Records (SGDR), with the full waveform data every 0.055 s (372 m along track). This gives us not only the finer spatial resolution, but also the opportunity to examine and reinterpret the waveform data in the knowledge that it is from a complex environment.

**3. Results**

**a. Analysis of RA-2 waveform data**

We show two examples of different waveform series for the same track at different time periods: 5$^{th}$ July 2006 on cycle 49 and 22$^{nd}$ November 2006 on cycle 53. Although Envisat follows almost the same track for the two transits (the repeat in cycle 53 being ~1500 m east of that in cycle 49), there is a marked difference in the observed waveforms. For cycle 53 (right-hand set of waveforms in Fig. 1) the echo returns are "Brown-like", whereas for cycle 49 (left-hand side) the waveforms show a complex structure with a significant power increase superimposed on the ocean-like returns. At the start of the sequence (waveform no. 1) this extra energy is in wavebins 105-115 and only just noticeable. By the 11$^{th}$ location (waveform no. 3) this "power excess" has migrated towards the front of the waveform trailing edge and is more marked. Waveform no. 5 shows this moving feature to have reached the leading edge of the waveform, where this additional component is three times the magnitude of the Brown-like signal, with waveform no. 7 showing the feature broadening and receding as the satellite leaves the island.

That these anomalies in cycle 49 waveforms have a common origin is attested by Fig. 2, which shows the gradual changes through the waveform locations marked in Fig. 1. The waveforms in Fig. 2 have been corrected for tracker movements as in [11] i.e. compensation for the necessary changes in the time origin of the reception window on-board the satellite. This display of the waveforms reveals two hyperbolae: one a bright target centred at 42.595°N (only in the lower plot), and the other corresponding to a power deficit caused by weak reflection from land, centred at 42.583°N. The size of these hyperbolae is determined by the altimeter's orbital and sampling parameters (see section 3b).

Such features have been noted before, for example, with a bright target such as a radar transponder [12], whilst negative features (i.e traces of localised weak signal) were demonstrated for cases of attenuation by small rain cells [13] and successfully modelled [14], [15]. The issue of small localised peaks in ocean backscatter ("sigma0 blooms") was first identified by *G. Hayne,* [pers. comm., 1996], with origin attributed variously to a sudden drop in winds, presence of a freshwater slick or the existence of biogenic surfactants. [11] catalogued many examples in the open ocean, demonstrating how the change in waveform shape sometimes led to problems for the waveform retracker. However, bright targets peaking ahead of the normal front of the waveform are due to reflecting surfaces above sea level [16].

**b. Modelling of waveform data**

To understand the causes of the hyperbolic feature observed, we developed a simple mathematical model encompassing Brown-like returns from the general oceanic background, with superposition of the effects of weakly-reflecting land and a small bright target. The model is described by:

$$\text{Waveform} = \text{Brown-like} + \sum_{i=1}^{n} A_i \exp\left(-\frac{1}{2}\frac{(\tau - \tau_i)^2}{\sigma_i^2}\right) \quad (1)$$

The simulation was done following [5] adapted to Envisat RA-2 specifications, with the second part of the equation modelling the *n* bright targets as narrow Gaussian peaks, where $\tau$ is the time (in units of waveform gates), and the $i^{th}$ feature has amplitude, $A_i$ (in picowatts), position (time delay), $\tau_i$ and width $\sigma_i$ (both in units of gates). The migration of the peak along the trailing edge can be modelled in a predictable way. Thus at time t the target is at gate $\tau_i$ given by the expression:

$$\tau_i \approx \alpha_0 + \frac{v^2}{c}\frac{(t-t_o)^2}{H_0} \cdot \frac{1}{gs} \quad (2)$$

where $\alpha_o$ is the value of the tracking point (in units of gates), $v$ is the satellite velocity, $c$ is the speed of light, $t$ is time along the track with $t_0$ being time of closest approach, $H_0$ is the effective satellite height (see Eq. A1 of [15]) and $gs$ is the gate spacing. Note that as the eccentricity of the hyperbola is close to unity, we have simplified the mathematical form in Eq. 2 to that of a parabola. The physical constants for the Envisat spacecraft and altimeter are $H_0$ = 695 km, $v$ = 6.7 km s$^{-1}$, $\alpha_o$ = 46 gates, and $gs$ = 3.125 ns. The number of bright targets was $n$ = 1 and wave conditions characterized by SWH = 0.5 m. The land contamination observed in waveforms sections in terms of a power deficit (clearly seen in the trailing edge area) has been also modelled considering a negative 'linear' contribution for land effects. Using these prescribed values, and involving reduced reflectance from land plus a small region of enhanced reflectance in the northern bay, the model generates a very similar set of waveforms (Fig. 3). We used a very bright target with a width of 3 bins. This approximately corresponds to a size of 140m on the sea, but this is dependant on the reflectivity of the target. Note that to simulate the pattern of waveforms there is no need to fix the absolute scaling to match the data shown in Fig. 2. Here, the parameters describing the land and bright target were chosen by trial and error, but their extraction through the application of fitting methods will be explored at a later date. *Synthetic Aperture Radar* (Envisat ASAR) images show the existence of intermittent calm water slicks in this region running along the area.

**4. Possible causes of intermittent feature**

We have examined waveform data from ascending track 128 over Pianosa for 46 passes during the Envisat mission, and found that on 30 of them (65%) the waveforms show similar behaviour to that observed at cycle 53 where the small section of the island and its flat topography merely serve to produce a weak loss of power in the signal (Fig. 2a). However, in one third of cases the hyperbolic signature of a localised bright target is found, and nearly always in the same place, approximately 2 km offshore in the Golfo della Botte (Fig. 2b). In this section we consider the various hypotheses suggested as the cause of this intermittent feature. First the bright target may not necessarily be at nadir, but any point perpendicular to the track at the apex of the hyperbola, provided that its height above sea level is sufficient to give the requisite time delay. Thus off-ranging to land could explain such a feature, however we know of no mechanism that explains the intermittency. Although the altimeter does not always follow exactly the same track, its longitudinal variations are of the order of 1500m (Fig. 4a) and uncorrelated with the presence of a bright target response. There is no obvious seasonality, as would be expected if changes in vegetation were responsible, and the idea of occasionally flooded swamps can be discounted because there is little rain on Pianosa. Accepting that the bright returns are from nadir, one further "land" explanation can be advanced: the revealing and concealment of large sand banks by the tide. However, the tides in the region are ~0.2m [17], and Envisat's sun-synchronous orbit aliases the main tides to very long periods.

Nadir backscatter from the sea surface varies inversely with the amount of small-scale waves, typically at wavelengths a few times that of the imaging radar. The $K_u$-band wavelength of RA-2 is 2.2 cm, but the hyperbolic features are also echoed in the S-band signal (wavelength = 9.4 cm) implying a reduction in sea surface roughness spanning a wide range of wavelengths. The strength of surface roughness is normally associated with wind speed, although wave conditions may play a part. Here we examine a number of sources of metocean data. First we look at wind speed information from the altimeter itself, taking the value near 42.9°N (Fig. 4b), which is far enough from the island and the bright target to guarantee a Brown-like return. An alternative source is the local meteorological records (at station S on Fig. 1), for which we show the wind speed and direction (Figs. 4c & 4d). Wind speed shows no clear connection with bright targets, and wind direction alone cannot provide a physical cause, although we note that wind from the southeast generally leads to low wave height (Fig. 4e), Swell (which may depend upon wind direction) is not an explanation, because the many surrounding islands limit the fetch in all directions.

A flotilla of small fishing boats cannot explain the signal, as this is a marine protected area, and even big ships cannot produce an echo that dominates the waveform after its leading edge [16]. The final explanation advanced was that of biogenic slicks of surfactants, generating a locally smooth surface [18]. However the occasions when a bright target is present do not coincide with the end of a bloom (see Fig. 4.f from MODIS data) when the phytoplankton are suffering nutrient stress. Thus, although it seems likely that the intermittent bright target is due to modulation of the sea surface roughness, a direct physical cause cannot be determined without the use of dedicated *in situ* measurements. The case studies in Fig. 1 (and corrected for in Fig. 2) show much less tracker window movement when the bright target is present; however an analysis of a fuller dataset shows the degree of tracker movement (Fig. 4g) does not depend on either the wind speed or the existence of a bright target.

## 5. Concluding remarks

In this study, we analyzed multiple series of Envisat RA-2 waveforms from around Pianosa, considering reflections from ocean, land and a regionalised bright target. We analysed forty-six repeat passes along ascending track 128. In nearly all cases, land is noticeable through a broad pattern of "power deficit" due to the lack of strong reflection from the sea. In 35% of cases a narrow hyperbolic bright target response is present in the Golfo della Botte on the northern side of the island. Although many explanations have been espoused, none of the available data on winds, waves or biological production show a good match to the bright target's existence. According to SAR data, calm waters are a common feature in this area, although neither the SAR nor the altimeter data reveal whether the cause is solely due to wind field or also biogenic material.

We developed a simple mathematical model able to replicate the features shown, proving that the moderately complex waveforms in a region surrounding a small island can be decomposed to enable marine altimetric applications right up to the coast. The modelling procedure presented in this paper might be utilized to better predict the aforementioned effects simultaneously through a set of waveforms and then subtract those effects in the retracking process. This process could be automated using techniques for fitting hyperbolae to waveform data [15], [16]. Thus the accuracy in the retrieval of geophysical parameters should improve especially in the coastal zone. This will help minimise differences between altimetry and tide gauges, and so help improve the fusion of sea surface height data from these two different datasets. To demonstrate this potential more fully will require integration with efforts to improve the altimeter corrections in the coastal zone, and validation for overpasses of islands with long-established tide gauges.

## 6. Acknowledgments

The authors are grateful to the many interested people at OSTST 2009 (Seattle) who offered various suggestions as to possible causes. We also wish to thank Donata Meneghello and Umberto Sassoli from Geographical Service of Regione Toscana (Italy) for having kindly supplied the high resolution coastline and DEM for Pianosa, the Meteo Service of Regione Toscana ([www.lamma.retetoscana.it](www.lamma.retetoscana.it)) for providing rain and wind data for the island, and Andrea Orlandi for his assistance in exploiting wind and wave data from their operational model. MODIS ocean colour data were obtained from NASA/GSFC. This work was done in the framework of the COASTALT project (N. 20698/07/I-LG) funded by ESA and was partially funded by the Spanish Research & Development Programme (N. CGL2008-04736). SGDR cycles were provided by ESA under Category-1 project id: 5785.

**Figure captions**

Figure 1: High resolution Digital Elevation Model of Pianosa (NW Mediterranean) with superimposed the exact location of Envisat ground track 128 at two different cycles. On the right (left) side, the four subplots show the waveforms at selected locations for cycle 53 (49). S marks the meteorological station.

Figure 2: Examples of Envisat RA-2 waveform data along track 128 over Pianosa: a) Cycle 53 showing the typical Brown-like return and b) Cycle 49 showing the complex structure resulting from a "power excess". White vertical lines are the southern and northern limits of Pianosa Island.

Figure 3: Example of a simulated waveform along track 128 obtained using the proposed model for RA-2 specifications.

Figure 4: Variations in environmental conditions for March 2006 to January 2007: a) Longitudinal position of altimeter transit at 42.6°N; b) Wind speed from altimeter; c) Wind speed and d) Direction from measurements at station S (see Fig. 1); e) Wave height from altimeter; f) Chlorophyll concentration from MODIS; g) Extremes of tracker movement on crossing island. Data are from the 9 passes for which data have been obtained from met station S, with the bold symbols indicating waveform series showing a hyperbola.

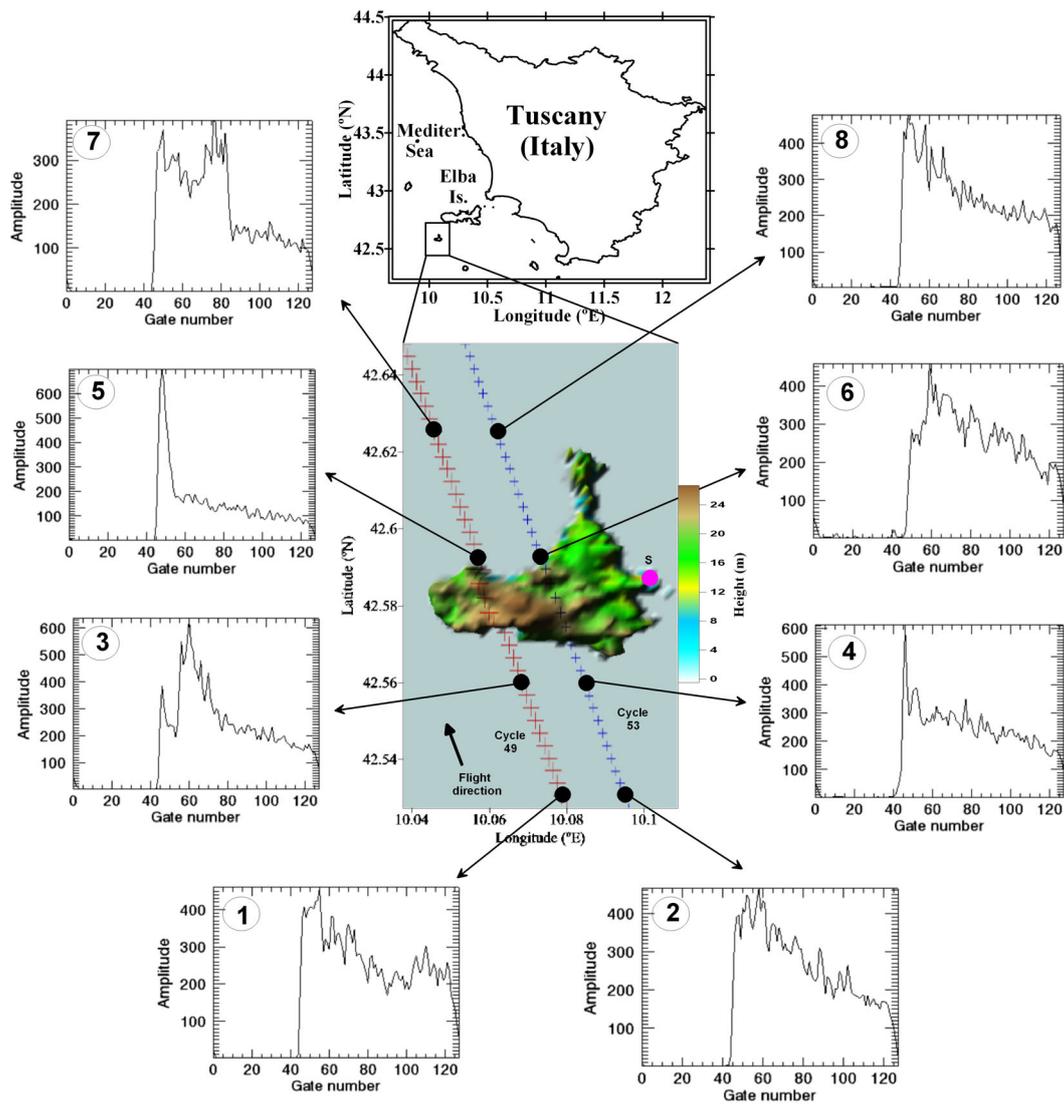

Figure 1: High resolution Digital Elevation Model of Pianosa (NW Mediterranean) with superimposed the exact location of Envisat ground track 128 at two different cycles. On the right (left) side, the four subplots show the waveforms at selected locations for cycle 53 (49). S marks the meteorological station.

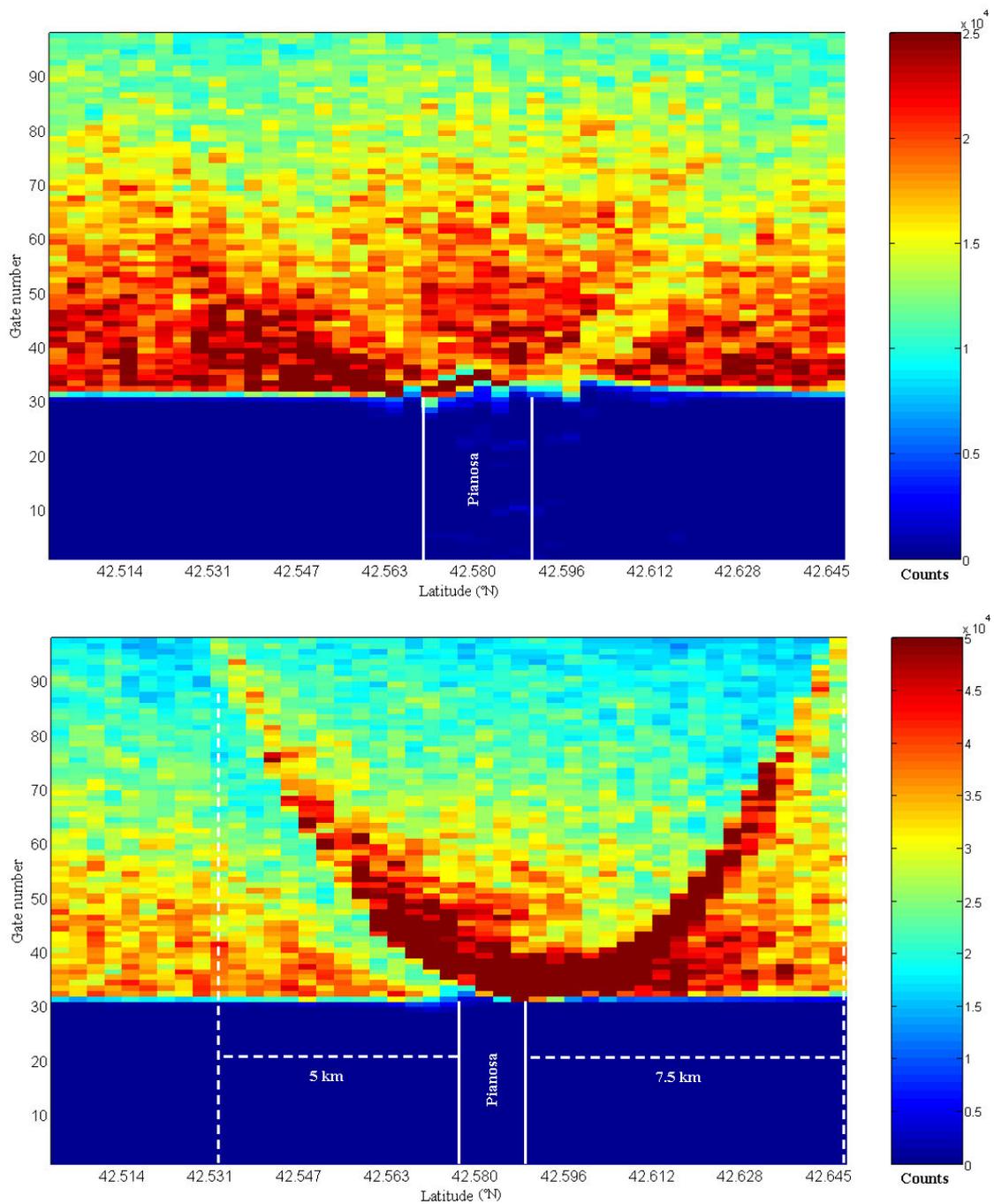

Figure 2: Examples of Envisat RA-2 waveform data along track 128 over Pianosa: a) Cycle 53 showing the typical Brown-like return and b) Cycle 49 showing the complex structure resulting from a "power excess". White vertical lines are the southern and northern limits of Pianosa Island.

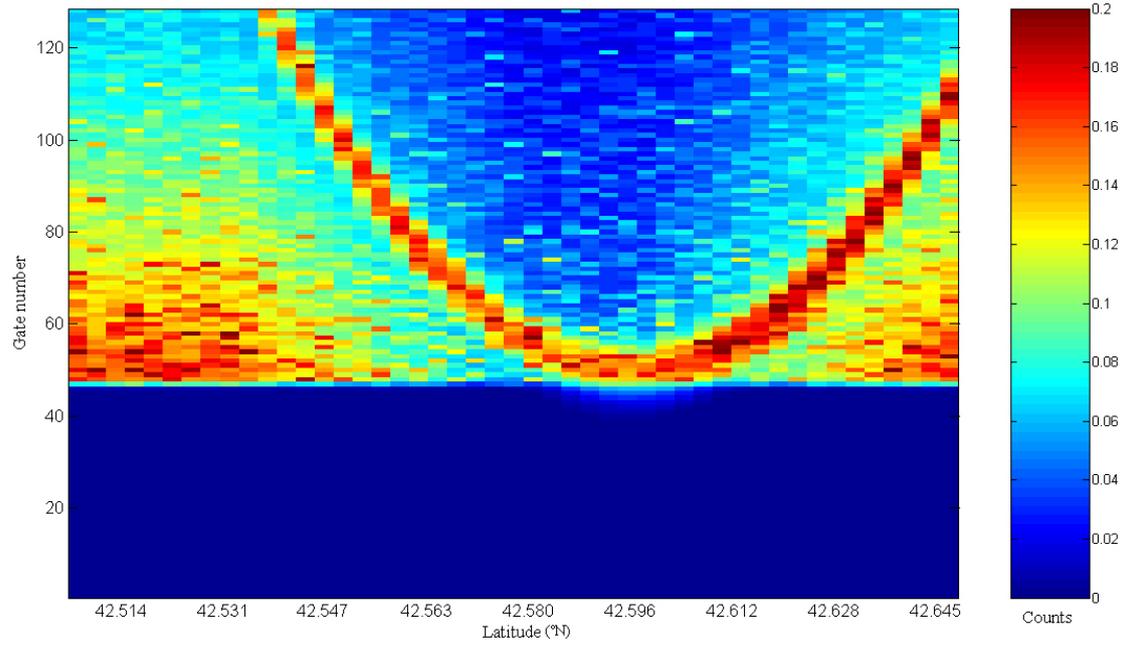

Figure 3: Example of a simulated waveform along track 128 obtained using the proposed model for RA-2 specifications.

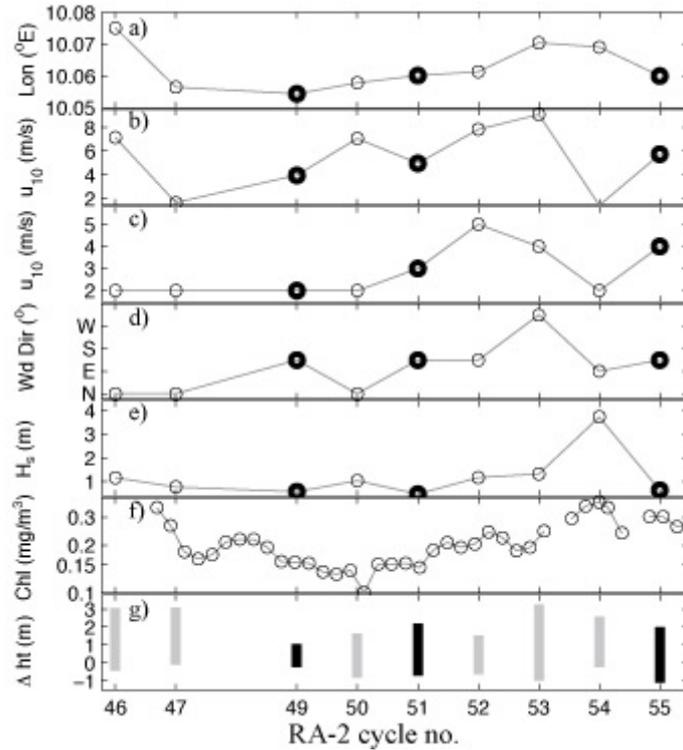

Figure 4: Variations in environmental conditions for March 2006 to January 2007: a) Longitudinal position of altimeter transit at 42.6°N; b) Wind speed from altimeter; c) Wind speed and d) Direction from measurements at station S (see Fig. 1); e) Wave height from altimeter; f) Chlorophyll concentration from MODIS; g) Extremes of tracker movement on crossing island. Data are from the 9 passes for which data have been obtained from met station S, with the bold symbols indicating waveform series showing a hyperbola.